\documentclass[11pt]{article}

\usepackage{amssymb,amsmath}
\usepackage[dvips]{graphicx}
\usepackage{epsfig}
\usepackage{cite}

\def\laq{~\raise 0.4ex\hbox{$<$}\kern -0.8em\lower 0.62
ex\hbox{$\sim$}~}
\def\gaq{~\raise 0.4ex\hbox{$>$}\kern -0.7em\lower 0.62
ex\hbox{$\sim$}~} \numberwithin{equation}{section}

\def\be{\begin{equation}}
\def\ee{\end{equation}}
\def\bea{\begin{eqnarray}}
\def\eea{\end{eqnarray}}

\def \6 {{}^{(6)}}
\def \5 {{}^{(5)}}
\def \4 {{}^{(4)}}

\begin{document}

\begin{titlepage}

\begin{center}

\textbf{Breaking the symmetry. \\
The first steps of a new way of thinking.}

\vspace*{1.5 cm}

\large{I. A. Sardella}

\normalsize \vspace{.2in} {\sl Seminario di Storia della Scienza,
Universit\`a degli Studi di Bari \\

Piazza Umberto I, 70121, Bari, Italy \\ e-mail:
eziosardella@libero.it}

\normalsize \vspace{.2in} {\sl PACS \\ 01.65.+g  - History of
Science
\\ 05.70.Fh  - Phase transitions: general studies \\
67.25.dj   - Superfluid transition and critical phenomena \\
11.15.Ex - Spontaneous breaking of gauge symmetries}

\vspace*{1.5 cm}
\begin{abstract} The concept of Spontaneous Symmetry Breaking
(SSB) represents a real breakthrough for present description of
fundamental interactions by means of gauge theories. Although the
underlying ideas were ancient, their formalization required a long
time, due to epistemological obstacles and technical difficulties.
In this paper, the main steps of SSB evolution are briefly
outlined, from the introduction of the order parameter in the
Thirties to the birth of the many-body theory at the end of the
Fifties. In this context, the contribute of the capital L.
Landau's works on phase transitions and quantum fluids, as well as
of the seminal ideas of F. London, is highlighted, and the
phenomenological approach in theoretical physics (whose features
are schematically underlined) is showed to be crucial in the
rising field of complex systems.

\end {abstract}

\end{center}

\end{titlepage}

\section{Introduction}

The role of Spontaneous Symmetry Breaking (SSB from now on) in
theoretical physics of the last fifty years can be hardly
exaggerated. In the Standard Model it represents the key concept
to achieve the formalization of the electroweak sector of the
theory on the base of gauge field symmetry. Moreover, the Higgs
mechanism, accounting for the mass of the intermediate vector
bosons, is a next decisive step towards the unification of forces
on the ground of symmetry principle \cite{BBCHN}. However, the
role of SSB cannot be limited in the area of particles physics;
because it represents the pivot of the building of the general
theory of phase transitions, it is obviously widely used in
ferromagnetism and ferroelectricity, as well as in superfluidity,
superconductivity, Bose-Einstein condensation phenomena \cite{HU}.
From a philosophical point of view, SSB has sometimes been
considered as the theoretical background for the understanding of
emergence phenomena, in the context of a non-reductionist
conception of the hierarchical structure of theories
\cite{A,Cast}.

Despite of relevance, the first steps of SSB have not yet received
from historical studies an adequate attention, apart from
reminiscences by physicists\footnote{Precious recollections are
available in the Archives for the History of Quantum Physics
(AHQP, Niels Bohr Library, American Institute of Physics, College
Park, Maryland). See in particular the interviews with: F. Bloch
(1964, 1968, 1981); P. W. Anderson (1987, 1999); W. Heisenberg
(1962, 1970); Y. Nambu (2004).} \cite{N,BRE,S} and not many
recollections by professional historians \cite{HBTW,BC}. As a
matter of fact, scholars are mainly devoted to the so-called
revolutionary periods of science, whereas SSB as a formalized
concept started after the quantum mechanics paradigm was
established (in Kuhnian words: in a \emph{normal} phase of
science). Moreover, it has to be evidenced the objective
difficulty of retracing a long and tortuous path, only recently
completed, which involves the interlacement of different search
fields (not yet recognized as autonomous at the time).

In this paper, an essential historical reconstruction is tempted.
Keeping in mind that SSB, perhaps more than other theoretical
issues, was the result of a collective work, there will be
mentioned only some of the essential contributes. Quite
conventionally, the end of the story can be fixed at the beginning
of Sixties. At that time, SSB idea acquired a general theoretical
meaning, since it was introduced in the framework of particle
models, coming from the original area of condensed matter physics.
After Heisenberg seminal ideas at the end of the Fifties, this
\emph{cross fertilization} was mainly due to Y. Nambu and G.
Jona-Lasinio papers about the analogy between the BCS theory of
superconductivity and the quantum field theory \cite{JL}. Much
more difficult to locate the beginning of the story. Maybe, it can
be placed at the end of XVIII century, with first Eulero's
considerations about a cylindrical rod which, subjected to an
increasing force, flexes in an apparently casual direction
\cite{E}; or, at the half of XIX century, with Jacobi's work about
a mass subjected to self-gravitation which, due to the rapid
rotation, changes its initial spherical form into ellipsoidal
\cite{J}. In these and other similar observations, it can be found
a symmetrical situation which is converted to a less symmetrical
one when a parameter exceeds a given value\cite{R}.

\section{Why such a long road to SSB?}

Now, a first question could be asked: why did SSB take such a long
time to emerge as a general theoretical concept? An answer could
be given paying attention to: 1) the philosophical conception
supporting SSB and epistemological obstacles (in the Bachelard's
sense \cite{B}) interposing to its acceptance; 2) the network of
mathematical and physical tools involved in a formalized SSB.

The main idea of SSB is clearly expressed by the inventors of the
term, Baker and Glashow \cite{BG}: it is plausible that the real
world complexity is not reflected in an equally intricate
fundamental theory. The idea of SSB tries to satisfy a
comprehensible desire of a scientist: the asymmetry (and
complexity) of physical phenomena does not imply that we cannot
use symmetry (and simplicity) in describing Nature at a
fundamental level. In mathematical terms: the complexity is not
present in equations; it arises only in solutions. In this
conception it can be found a sort of reconciliation between two
important ways of thinking in Occidental culture \cite{H}: the
idea that the laws of Nature are obscure and inaccessible, because
of complexity of physical phenomena and, on the other side, the
representation of Nature as an open, intelligible
book\footnote{The first position is well expressed by the famous
Eraclito's sentence: ''Nature loves to hide''; the second is
poetically rendered by the verses of Goethe: ''Seize, then, with
no delay, the sacred mystery in broad daylight''\cite{GOE}.}. It
can be thought that an epistemological obstacle may be arose from
the philosophical difficulty to conceive a disagreement between
the deep essence of reality (the laws of Nature) and the
manifestation of it (the physical phenomena). This can be seen
more clearly reminding the reflections of Pierre Curie, the first
one at the end of XIX century to stress the importance of
asymmetry in physics\footnote{The deep interest in symmetry
questions by Pierre Curie, hair of the prestigious French school
of crystallography, led him first to the discover of piezoelectric
effect and then to an acute analysis of the reflection properties
of electromagnetic fields.}. In 1894 he stated the following laws:
1) Why a phenomenon may appear, it is necessary that certain
elements of symmetry are missing: \emph{c'est la dissymm\'{e}trie
qui cr\'{e}e le ph\'{e}nom\`{e}ne}; 2) The symmetry of the causes
must be preserved in effects; 3) The asymmetry of a phenomenon
must arise from asymmetrical causes \cite{C}.

In the first law Curie made an important step towards SSB. Indeed,
he underlined the connection between the occurrence of a
phenomenon and the generation of asymmetry conditions compatible
with it\footnote{Pierre Curie expressed his first law in terms of
group theory. He applied the law to several physical situations,
such as the Wiedemann effect, noting that it can give only a
necessary condition for the occurrence of a phenomenon.}. However,
in the last two laws, Curie assumed that causes and effects must
have the same degree of symmetry, according to Leibniz principle
of sufficient reason. That was really a strong paradigm, in
scientific community. As evidence of this, L. Radicati
mentions\cite{R} that Curie did not interpret the experiment of
his colleague B\'{e}nard in 1900 as a falsification of his laws.
In B\'{e}nard's experiment, a liquid in a beaker was uniformly
heated from below. As soon as the vertical temperature gradient
reached a critical value (later measured by Rayleigh), a regular
pattern of convection cells appeared, breaking the symmetry in
horizontal planes \cite{BEN}. From considerations dating back to
Euler\cite{E}, this kind of phenomena could be explained through
subtle asymmetric sources existing in causes and manifesting in
macroscopic evidence in effects; however, it seemed difficult to
accept, in opposition to a generally recognized \emph{continuity
principle}, that such effects could manifest only above a definite
value of a parameter. A mature awareness of the typical
consequences of non-linearity occurred only much later.

In 1950 the American mathematician G. Birkhoff dedicated the first
chapter of his book \emph{Hydrodynamics}\cite{BI} to the so called
\emph{paradoxes} (in which he mentioned B\'{e}nard's experiment).
He ironically opposed the established beliefs stating that
''Nature has desire of symmetry at the same way that abhors the
vacuum'' and explained macroscopic manifestations of asymmetries
on the ground that ''although the symmetric causes should produce
symmetrical effects, nearly symmetrical causes not necessarily
produce almost symmetrical effects; a symmetric problem does not
necessarily have a \emph{stable} symmetric
solution''\footnote{Birkhoff imputed to physicists and engineers a
lack of mathematical rigor in interpretation of hydrodynamic
phenomena. He reminded the Oseen result of the \emph{limit
paradox}: in systems of differential equations, the presence of
arbitrary small terms of high order can completely change the
aspect of the solutions.}.

In more recent years, G. Jona-Lasinio recalled the resistance
expressed from particle physics community towards the acceptation
of SSB concept. After defining SSB in condensed matter as ''the
situation occurring when the lowest energy state of a system can
have a lower symmetry than the forces acting among its
constituents and on the system as a whole'', he remarked that ''to
appreciate the innovative character of this concept in particle
physics one should consider the strict dogmas which constituted
the foundation of relativistic quantum field theory in the late
Fifties. One of the dogmas stated that the lowest energy state,
the vacuum, should not possess observable physical properties, and
all the symmetries of the theory, implemented by unitary
operators, should leave it invariant''\cite{JL}.

Apart from the overcoming of epistemological barriers, the
elaboration of SSB idea required the development of a quantity of
theoretical tools.

In order to highlight this point, it is useful to sketch a modern
possible definition of SSB. The framework in which this idea can
be found is that of a theory expressed by non linear equations,
invariant for transformations by a continuous symmetry
group\footnote{It is possible to have SSB with a discrete symmetry
group, such as in the Ising model, but continuous symmetries are
mainly involved in gauge field description of fundamental
interactions.}, with a degenerate vacuum structure (i.e. set of
physically equivalent states, transforming according to the
symmetry group). In thermodynamic limit (number of degrees of
freedom which tends to infinity), by varying some parameters a
bifurcation point can occur in which stable solutions have a lower
symmetry than the lagrangian. The \emph{jump} to the less
symmetrical states can be due to fluctuations. The breaking of
continuous symmetries are manifested by the existence of
collective excitations, whose quanta are the so-called
\emph{Goldstone bosons}, massless, corresponding to the group
generators in the broken symmetry directions.

As it can be seen, SSB is a confluence of sophisticated
theoretical concepts coming from linear algebra and group theory,
phase transitions, theory of fluctuations and collective
excitations. All these theories were laboriously developed over 50
years, on the basis of the pillars of statistical mechanics and
quantum mechanics, in a process that led to the birth of the
so-called many-body theory.

\section{Measuring the order}

Taking into account the above considerations, a possible schematic
reconstruction of the nodal points of historical process driving
to SSB can be exposed as follows. At the end of XIX century,
capital works by S. Lie and F. Engel about the invariance of
equations in symmetry transformations stressed the importance of
symmetry considerations in the laws of physics \cite{SL}. On the
other hand, Pierre Curie underlined the role of asymmetry in
physical phenomena. It was just the tension between these two
instances which became the natural background for a careful
analysis of phenomena, like ferromagnetism, deeply involved in
symmetry questions. Ferromagnetism had an essential role inasmuch
it allowed, in a context sufficiently explored in its macroscopic
features (among others, by P. Curie himself, whilst microscopic
details were to remain mysterious at least until the famous
Heisenberg's work about the quantum exchange interactions in
1928), the easier comprehension of some of the main elements of
complex systems: asymmetric ground states, relevance of
fluctuations at critical point, collective excitations.

At the beginning of last century, a very important step was made
by the French physicist Pierre Weiss. First, because he focused
the attention on ferromagnetism in crystals (in this way, he was
led to emphasize the role of symmetry questions in such a
phenomenon\footnote{With respect to this point, it was very
important the availability of high quality crystals. Weiss managed
to seize pyrrhotite crystals from Brazil perfectly homogeneous
over long distances\cite{HBTW1}.}); more important, by virtue of
his conception of ferromagnetism as a \emph{cooperative
phenomenon}. Indeed, Weiss deduced the relationship between
magnetic susceptibility and temperature by means of the
introduction of a \emph{molecular field}\footnote{Suggested to him
by the internal pressure parameter in van der Waals equation of
real fluids.}, in order to take into account the collective action
of the rest of the system on a given elementary magnetic dipole
\cite{PW}. The fundamental assumption of such a \emph{mean-field}
approach was that of proportionality between the molecular field
and the magnetization, that is to say, the degree of asymmetry
existing in the system in the ferromagnetic phase. So, Weiss was
able to obtain a self-consistent equation to calculate the
magnetization and identify a definite temperature at which the
transition occurs between the paramagnetic and the ferromagnetic
phase\footnote{In 1910, the transition temperature was named
\emph{Curie point}, on a proposal by P. Weiss and K. Onnes, even
if then French physicist localized only an interval of
transition.}. This kind of approach was phenomenological because
the molecular field was not really deduced from molecular
interactions.

In the first decades of century, the analysis of ferromagnetic
systems intertwined with the issues coming from a still lacking
theory of phase transitions. In a pretty obscure paper in 1914 the
Dutch physicists Ornstein and Zernike stressed the importance of
density fluctuations in phase transitions deducing the phenomenon
of critical opalescence from the existence of density correlations
which increase indefinitely in range as the critical point is
approached\cite{OZ}. That was really a crucial step for phase
transition understanding, since it allowed to go beyond the
statistical mechanics assumption of neglecting fluctuations in
favor of the average values.

At the end of Twenties, the mean field technique adopted in
ferromagnetism proved to be very useful in theoretical approach to
the so-called \emph{superlattices}, a new line of research in the
field of crystallography. At that time, mainly due to constant
advances in resistivity measures and x-ray spectroscopy
technology, a precise analysis of the lattice structure of binary
alloys had become possible. As a clear experimental result, at low
temperature the atoms of two different metals were found to be
strictly ordered in a regular pattern (eg.: each atom of one metal
surrounded by atoms of the other metal, in a body-centered cubic
lattice), whilst at high temperature they appeared to be
accidentally distributed among the sites of the
lattice\footnote{The first indications for such behavior were due
to Tammann in 1919.}. Through a series of investigations mainly
due to Gorsky in Leningrad, Dehlinger in Stuttgart and the group
of Borelius, Linde and Johansson in Stockholm, it became evident
that it was involved a transition with no latent heat, occurring
in a fairly small temperature range and accompanied by large
specific heat and electric resistance\cite{GBJLD}. In order to
describe a transition within two solid phases, these physicists
realized that a new thermodynamic parameter was needed, whose aim
was to measure the degree of order in the lattice due to the
atomic distribution. In 1934 the British physicists W. L. Bragg
and E. J. Williams defined the concept in a profitable way to
build a mean field theory of superlattices\footnote{Bragg and
Williams defined the parameter $S$, which they called \emph{degree
of order}, in the following way: $S=\frac{p-r}{1-r}$, where $p$ is
the actual probability that a given position in the lattice is
occupied by the atom required for complete order; $r$ is the value
of $p$ in case of complete disorder. When order is complete, $p=1$
and $S=1$; at transition point, when order disappears, $p=r$ and
$S=0$. However, they recognized the birthright of Gorsky in 1928
in identifying a degree of order and applying the mean field
method to the alloys.}. Under the crucial assumption of the linear
dependence of the potential energy involved in an atomic
interchange from the order parameter, they obtained a
self-condition equation which gave a sharp transition point
\cite{BW}. In a series of three papers, they clarified the
relation between the order parameter and the presence of long
range order in the lattice. Successively, Bethe\cite{BE} and
Peierls\cite{PA} introduced a new parameter (unlike the previous
one, persisting above the transition point) to take account for
short range correlations, neglected by Bragg and Williams, and
improve the approximations. In the following debate, interesting
issues related to the evaluation of long and short range order in
mean field theories were addressed\cite{KI}.

As a matter of fact, the order parameter was conceived as an
instrumental tool for mean field theories in order-disorder
transitions. However, its introduction was also needed to identify
continuous transitions even in situations where the presence of
"order" was less obvious. In the early Thirties, the Dutch
physicists Gorter and Casimir, with a phenomenological
approach\cite{GC}, applied thermodynamics on
superconductivity\footnote{The thermodynamic approach was
supported in Netherlands by Ehrenfest and Rutgers even before the
fundamental Meissner and Ochsenfeld's experiment in 1933 that,
showing the phenomenon of screening of the magnetic field from the
interior of the metal, proved the reversibility of the
superconducting state and launched the new paradigm of
superconductor as a perfect diamagnetic material.}. Adopting
Kronig's hypothesis of superconductive phase as a coexistence of a
gas of free electrons with a sort of a lattice of electrons, they
considered the transition from superconducting to normal phase as
a phenomenon of evaporation of electrons. Defining a so-called
"internal parameter" as the fraction of the electrons being in the
lattice, they identified the transition temperature as the point
in which that parameter vanishes. More important, they used the
concept as a way to affirm, in a lively exchange of views with von
Laue and Justi, an essential feature of a continuous transition:
the transformation of one phase into another without metastable
states \cite{GOR}.

\section {The Landau's theory of phase transitions}

In 1937, Landau's work about the theory of phase transitions
\cite{LA1} responded to the need for clarification of key
questions relating mainly to the concept of continuous transition.
Suggested by Keesom's observation of a peak in the specific heat
of $^{4}$He at the so-called $\lambda$-point (i.e. the transition
temperature between the superfluid and the normal phase), the
Ehrenfest classification of phase transitions based on
discontinuity of the derivatives of thermodynamic potentials
showed the possibility of a general description but did not clear
crucial questions such as the conditions of their occurrence, the
physical mechanism driving to discontinuities, the possibility of
metastable states\cite{JA}. The fundamental Landau's belief was
that such transitions can be understood in terms of a change in
the symmetry properties of the stable states of the system,
through a mechanism of spontaneous symmetry breaking. That
conception implies that in a continuous transition an order
parameter is always involved, and a proper description can only be
given in terms of its variations\footnote{In comparison with
previous conceptions, the order parameter assumed with Landau a
more crucial role, because it became a general measure of the
breaking of symmetry in stable states. This meaning was the key to
its extension to more abstract fields than crystallography.}.
Landau referred to the principle: \emph{a given symmetry element
is either present or absent} to explain the transition as a
process of \emph{abrupt} reduction of symmetry\footnote{In
algebraic terms, the symmetry group of one phase (in general, the
\emph{disordered} or symmetric phase) is required to admit as a
subgroup the one of the other phase (\emph{ordered}, or asymmetric
phase).} in which, nevertheless, below the critical temperature
the order parameter starts to increase \emph{continuously} from
zero to non-zero values\footnote{In the Thirties, a debated
question was the possibility of a continuous transition between
the solid and the liquid phase. Y. Frenkel expressed the
conviction, based on supposed structural similarities between the
two phases, that elements of disorder can gradually appear in the
crystal lattice driving to a continuous transition to the liquid
phase. L. Landau strongly objected to this idea on the base of his
principle.}. Therefore, L. Landau assumed that the free energy
$\Phi$ could be expanded in series of the order parameter $\eta$,
near the critical point: $\Phi=\Phi_{0} + A \, \eta^{2}+ B \,
\eta^{4}+...$ (where only the even powers of $\eta$ are considered
for symmetry reasons and the coefficients $A$, $B$... are supposed
to be functions of the other thermodynamic variables of the
system\footnote{In\cite{LA1} the order parameter $\eta$ is first
deduced from the expression of the distribution probability of the
atoms of a crystal written in terms of the irreducible
representations of the its symmetry group; however, in
Landau-Lifshitz volume dedicated to Statistical Mechanics the free
energy expansion precedes the discussion on the distribution
probability, which indicates the intention to give it a more
general meaning.}). The value of $\eta$, at a given temperature,
can be obtained through the minimization of $\Phi$. Under general
assumption about the signs of the coefficients, the variation of
the order parameter allows to reproduce the phase transition, that
is the discontinuity in derivatives of $\Phi$.

The real Landau's breakthrough was the identification of a general
mechanism for SSB, in principle regardless of particular area of
application. Moreover, in the framework of the theory was easy to
show a very important feature of complex systems near the critical
point: the \emph{universal behavior}, due to the presence of long
wavelength fluctuations which cause the lack of relevance of
microscopic details.

Landau's theory had a relevant impact to Soviet scholars
\cite{PO}, but in Occidental countries in the early times it was
somewhat underestimated, for several reasons. Primarily, due to
problems of communication during the war; second, for mathematical
doubts about the convergence of the free energy series \cite{ST}.
Further, it should be reminded that it was still an open general
question the real possibility to reproduce the singular behavior
of thermodynamic potentials in a critical point by means of
analytic functions \cite{JDB}.

In 1944, after the Onsager's exact solution of two dimensional
Ising model \cite{ONS}, it became clear that phase transitions
could be captured by statistical mechanics methods, but also, that
Landau's theory suffered from the limitation of any mean field
approach in reproducing accurately the behavior of the system just
near the transition. As a result, its physical content was in some
extent obscured. In more recent years the theory was widely put
into great consideration: in 1984 Anderson defined Landau's
statement about the impossibility to change symmetry gradually as
the First Theorem of solid-state physics \cite{AN}.

\section{Collective excitations}

Goldstone theorem (1961) asserts that SSB of a continuous symmetry
group implies the existence of massless bosons due to collective
excitation in the direction of the breaking of symmetry \cite{GO}.
Since Debye's work in 1912 in which wave excitations in crystals
were considered in continuous limit (in that case, involving SSB
in rotational and translational symmetry), Goldstone's fundamental
issue was (unconsciously) anticipated from several condensed
matter results. Undoubtedly, an important step was Bloch's paper
in 1930 on ferromagnetism \cite{BL}. He studied the behavior of a
ferromagnetic material at very low temperature, showing that, at
the same way as Debye's propagation of elastic perturbations in
crystals, the deviations from perfect order (i.e. complete
alignment of spins) occur through propagation of spin waves. Of
these collective excitations, Bloch obtained the continuous
energetic spectrum at low frequency. Afterward, Soviet physicists
applied the Dirac's method of II quantization to elastic and
magnetic oscillation fields by introducing \emph{phonons} (Tamm,
1930) and \emph{magnons} (Pomeranchuk, 1944).

As a matter of fact, the collective behavior had a very important
role in Soviet approach to condensed matter physics\cite{KO1}. In
the Twenties, in the lively debate between the proponents of the
opposite models of conductivity based on completely free and quite
bounded electrons, Y. Frenkel\footnote {Y. Frenkel was the head of
theoretical division in the Physico-Technical Institute in
Lenigrad. L. Landau worked there, before establishing his famous
school in Kharkov.} proposed an intermediate model based on
collective sharing of electrons by the nuclei of the metal and,
later, introduced a quantum of collective excitation called
\emph{exciton}. Because of the lack of a satisfactory
formalization of his theory and of some redundancy of political
metaphors \cite{KO1}, Frenkel's ideas were not widely considered
in scientific occidental community. However, that line of research
was prosecuted, on one hand, by the the works of Bohm and Pines on
plasma in the Fifties \cite{KO2}; on the other hand, by the
fundamental Landau's work on superfluidity in 1941.

During the war years, large efforts of scientific community were
aimed at the great puzzles of \emph{quantum fluids}. The interest
was stimulated by the very unusual experimental behavior of
superconductors and superfluids that could be fraught with
technological implications, as well as by more theoretical
interest in suspected macroscopic quantum aspects and variety of
possible applications, such as atomic nuclei, neutron stars,
plasma. However, as soon it became clear, a shift was needed in
theoretical approach: from single-particle to many-particle models
in which, in principle, the interactions could not be neglected.

A very bold approach to superfluidity was tempted in 1938 by the
German physicist Fritz London, and prosecuted by the Hungarian L.
Tisza, colleague of Landau in Kharkov. Starting from the alleged
knowledge that the atoms of $^{4}$He are bosons, he observed that
the experimental measures of $\lambda$-point were not so far from
the Einstein estimation of the transition temperature in his
debated paper in 1925 about the \emph{Bose-Einstein condensation}
(BEC). Therefore, he proposed BEC as a microscopic mechanism for
superfluidity \cite{LO}. Assuming that at temperatures below
$\lambda$-point a finite fraction of atoms can be found in the
ground state of zero momentum (\emph{Bose-Einstein condensate}),
London could derive the zero viscosity of the superfluid part of
helium. The London idea was that it is possible to attribute a
single macroscopic wave-function to all the particles in the
condensate, which therefore moves coherently without friction. The
proposal was marked by controversial points: BEC was in doubt for
reasons of mathematical rigor\footnote{Among others, Uhlenbeck was
not sure, until 1937, of correctness of Einstein result. That
negatively influenced the acceptation of theory.} and, above all,
it referred to an ideal gas and it was not known in advance what
could be the effect of the interactions. However, London's view
was going to open a fundamental search line \cite{GRI}. From the
point of view of SSB history, it can be considered an essential
step, as it was put the focus on different, more abstract, kinds
of symmetry changes in a phase transition. Indeed, Bose Einstein
condensation involved an order in the space of momenta, rather
then in the configurational space as usual\footnote{In his
celebrated text \emph{Superfluids} in 1950, F. London gave reasons
to affirm that helium at zero pressure cannot maintain any
structure ordered in space; therefore, he proposed an alternative
form of order to justify the $\lambda$ transition.}.

The theory of London proposed a mechanism but was not able to
provide quantitative predictions. New technical, as well as
conceptual tools were required. Landau rejected London's approach
of BEC: in Soviet physics vision, the liquids were considered much
more similar to solids than to ideal gases \cite{KO1}. In that
conception, collective excitations were destined to assume an
increasing relevant role.

As a matter of fact, Landau's paper in 1941\cite{LA2} was really
decisive to introduce a new paradigm in condensed matter physics.
It can be stated in the following way: \emph{for any system with
strongly interacting particles it must exist a phenomenological
description in terms of an ideal gas of fictitious particles (or
quasiparticles), describing the collective excitations.} In this
framework, Landau gave a description of the very unusual behavior
of helium in the superfluid phase. In a fluid flowing through a
capillary the presence of viscosity, due to the kinetic energy
loss for frictions with the walls of the capillary and inside the
liquid itself, is manifested by the appearance of internal motions
in the form of elementary excitations. These can be seen as
longitudinal density waves (\emph{phonons}) and vortex excitations
(\emph{rotons}). Therefore, at finite temperature below
$\lambda$-point, helium liquid was considered by Landau to consist
of a gas of phonons and rotons in a superfluid background. The
quantitative character of the theory was assured by the assumption
of a particular phenomenological form of the energy spectrum of
the elementary excitations, in which some parameters had to be
fixed according to experimental results. In this way, Landau
predicted a critical velocity of the helium liquid (above which
superfluidity disappears owing to the excitations of phonons and
rotons) and the famous \emph{second sound}, a collective
excitation consisting of temperature oscillations.

\section{Complex is better than real}

The main difference between Landau and London's approaches to
superfluidity can be traced back in the relevance assigned, for
the first theory, to the properties of the low energy spectrum of
excitations; for the second, to the effect on the ground state of
the symmetry properties of wave-functions. The reconciliation
between these two aspects came in 1947 from a pretty
unacknowledged paper\footnote{In spite of its relevance,
Bogolubov's paper was not quoted in the important subsequent works
by Yang, Lee, Feynman and probably it was not known by London
himself. The difficulty of communication was surely due to the war
years climax, but Griffin also mentions the question of the still
not fully assimilated methods of field theory and the presence of
approximations with non-conserved number of particles in
Bogolubov's work \cite{GRI}.} by the Soviet physicist N. Bogolubov
\cite{BOG}. Adopting the field theory method of II quantization he
succeeded in demonstrating that BEC is not much altered by the
presence of weak interactions and that the presence in the state
of zero momentum of a finite fraction of atoms implies a spectrum
of low energy excitations very similar to that assumed by Landau.
Bogolubov's contribution was recognized much later as the first
step towards a microscopic theory of quantum fluids. However, at
the time the phenomenological approach again proved to be
extremely fruitful, in dealing with superconductivity.

Since the late Thirties, the most reliable description of
superconductive behavior (in case of weak magnetic fields) was
based on a set of phenomenological equations proposed in 1935 by
the two brothers Fritz and Heinz London  \cite{LOLO}. In order to
provide the behavior of perfect diamagnetism in a superconductor,
they assumed the principle that the superconductive density
currents $\textbf{J}_{s}$ are always determined by the local
magnetic field, according to the relation:
$\textbf{J}_{s}=-\frac{e^{2}}{mc}\, n_{s} \, \textbf{A}$, where
$\textbf{A}$ is the potential vector; $n_{s}$ is the concentration
of the superconducting electrons; $m$ and $e$ are mass and charge
of electron\footnote{London's equation can be considered as a sort
of magnetic Ohm's law for superconductors, in which the vector
potential replaces the electric field.}. In the concluding remarks
of the paper, the Londons tried to justify their phenomenological
relation from quantum mechanics considerations, starting from the
well known Gordon equation of the quantum current in presence of a
magnetic field: $\textbf{J}=\frac{he}{4\pi i m} \, (\psi \, {\bf
\nabla} \, \psi^{*}-\psi^{*}\, {\bf \nabla}
\,\psi)-\frac{e^{2}}{mc}\,\psi\psi^{*}\,\textbf{A}$, where $\psi$
is the wave-function of a single electron. In case of normal
conductors without magnetic field the total current is zero,
because summing over all the electron currents the first term
vanishes for symmetry reasons. In case of superconductors, the
Londons assumed the existence of an \emph{energy gap} (due to not
well specified electron coupling) between the ground state and the
first energy levels of the internal motions. This means that, with
sufficiently weak magnetic field, the electron wave-functions
remain essentially unperturbed\footnote{The property of
\emph{rigidity} of the ground state wave-function, supposed by F.
London for superconductors, was later recognized by P. W. Anderson
as one of the main macroscopic features of SSB\cite{AN}.}. Thus,
summing over all the electron currents the first term is always
vanishing and, interpreting $\sum\,\psi\psi^{*}$ as $n_{s}$, the
London relation results.

In 1950, in a fundamental paper \cite{GILA}, L. Landau and V.
Ginzburg elaborated a theoretical structure to explain the
superconductive behavior and deduce the London equation, through
an appropriate application of the Landau's theory of phase
transition. The starting point was that, having assumed the
existence of a continuous transition between the superconductive
and the normal phase, an order parameter must exist. Now, it is
clearly possible to develop a phenomenological theory without
establishing exactly the correspondence of any parameter to
physical quantities at a microscopic level; however, the theory
has to be structured so that such a connection could be eventually
found, in a \emph{first principle} approach. Therefore they
considered the order parameter (denoted with the symbol $\Psi$) as
a sort of an \emph{effective wave-function} of the superconductive
electrons which takes into account the overall behavior of the
particles. In so doing, they give to the order parameter a new
fundamental development assuming it as a \emph{complex function}.
The connection with microscopic quantum mechanics was proposed to
be realized through the \emph{density matrix} $\rho$, a
statistical object introduced in the early Twenties independently
by Landau and von Neumann as a quantum analogue of the phase space
probability measure. Ginzburg and Landau considered the
one-particle density matrix
$\rho_{1}\,(\textbf{r},\textbf{r}')=\int
\psi^{*}(\textbf{r},\textbf{r}'_{i})\,
\psi(\textbf{r}',\textbf{r}'_{i}) \,\textbf{dr}'_{i}$. In this
expression, $\psi(\textbf{r},\textbf{r}'_{i})$ is the true
wave-function of the $N$ electrons in the metal, depending on the
coordinates of all electrons $\textbf{r}_{i}\,(i=1,2,\dots, N)$;
the integral is extended to the coordinates $\textbf{r}'_{i}$ of
all the electrons except the one considered; the density matrix
refers to the two points given by the coordinates $\textbf{r}$ and
$\textbf{r}'$. Below the transition point, the order parameter
$\Psi$ must assume a value different from zero. In Landau and
Ginzburg view that implies, like in a ferromagnet, the existence
of long-range correlations, which is expressed by the condition:
$\lim_{\textbf{r}-\textbf{r}'\rightarrow
\infty}\rho_{1}\,(\textbf{r}, \textbf{r}')\neq 0$. Thus, to
establish the relation of electron wave-functions with the order
parameter they supposed to be valid the factorization:
$\lim_{\textbf{r}-\textbf{r}'\rightarrow
\infty}\rho_{1}\,(\textbf{r}, \textbf{r}')=\Psi^{*}(\textbf{r})\,
\Psi(\textbf{r}')$. Of coarse, an essential requirement of a
phenomenological theory is the accountability of previous
empirical laws (in this case, the London equations); in order to
obtain that result Ginzburg and Landau imposed the normalization
condition $|\Psi|^{2}=n_{s}$.

The theory was built in such a way to preserve the invariance for
global variations of the phase of the order parameter (the
so-called $U(1)$ gauge symmetry): to the expansion of free energy
in series of $|\Psi|^{2}$ was added the gradient term
$\frac{1}{2m} \, |-i\hbar \,
{\bf\nabla}\,\Psi-\frac{e}{c}\,\textbf{A}\,\Psi|^{2}$ to take into
account the interaction with the magnetic field by means of the
quantum coupling rule. Minimizing the free energy respect to
$\Psi^{*}$ and $\textbf{A}$, Ginzburg and Landau derived a set of
differential equations, whose solutions, giving the values of the
order parameter and the potential vector, allow to calculate the
superconductive current. In case of weak magnetic fields, the
London equation can be derived under the assumption that the zero
field constant solution $\Psi$ in not perturbed and so ${\bf
\nabla}\,\Psi=0$ can be placed in the expression of the current.

Landau and Ginzburg theory allowed a satisfying quantitative
description of superconductivity, which was valid for magnetic
fields of any intensity. In 1959, the Soviet physicist Gor'kov
\cite{GORK} was able to derive, in the neighborhood of the
critical point, the Ginzburg and Landau equations from the BCS
(Baarden, Cooper, Schrieffer) microscopic theory\footnote{In order
to get this result, the charge of the electron $e$ was replaced
with $2e$, which refers to the charge of the Cooper pairs.}.
However, apart of the particular context of application, the main
feature that acquired later enormous relevance was the extremely
flexible structure of the theory, useful for the formalization of
different theoretical fields\cite{ARKR}, and the highly
descriptive effectiveness of a SSB mechanism in which a complex
function was used as an order parameter. Indeed, in the next
evolution of SSB concept, a very important role was assumed by the
order parameter phase.

\section{Marching towards microscopic theories}

Starting from the Fifties, the aim of a large part of condensed
matter physicists became the search of a microscopic justification
of the phenomenological theories. First in superfluidity, then in
superconductivity, the key was found in Bose-Einstein condensation
and the symmetry properties of boson wave-functions. A preliminary
essential question was to be sure that BEC really occurred in
helium. Paradoxically, the technical way to answer the question
came just from the consideration of long range order correlations
in the framework of density matrix method, despite Landau was not
convinced of BEC occurrence.

In 1951, in Penrose seminal paper\cite{PEN} the existence of long
range correlations in the configuration space below the
$\lambda$-point (due to the extension of de Broglie wavelenghts of
the particles over the average interatomic distance) was expressed
in terms of density matrix and showed to be equivalent, according
to London's suggestion, to the condition of concentration of
particles in momentum space. Moreover, the factorization of
density matrix (which was previously proposed by Ginzburg and
Landau) was formally justified\footnote{In his paper R. Penrose
did not cited the work of Landau and Ginzburg. As a matter of
fact, the Ginzburg and Landau factorization was shown to be wrong
in the context of superconductivity, because of the fermionic
wave-functions involved\cite{YA}. In that case, a two-particle
density matrix must be considered. Anyway, the phenomenological
theory showed to be effective beyond the physical interpretation
of the order parameter.} and the function $\Psi$ recognized as the
wave-function of the single particle in the condensate state. In
such a way, the hydrodynamic behavior of superfluids was related
to SSB mechanism: the velocity of superfluid current resulted to
be connected to the gradient of the phase of $\Psi$. This means
that, although the laws of motions are gauge invariant, in a
superfluid at constant velocity the phase of the wave-function of
the particles in the ground state are fixed at an arbitrary but
coherent value for the whole condensate.

The work of Penrose was the first of a series of papers in which
Onsager-Penrose\cite{ONPE} and Yang\cite{YA} established strict
mathematical criteria to be satisfied in case of BEC (all based on
density matrix behavior below $\lambda$-point) which were denoted
as the \emph{Off Diagonal Long Range Order} condition (ODLRO).

However, an essential element was still needed for a microscopic
approach to superfluidity: a reliable ground state wave-function
to test the ODLRO criterion. This came out from the prominent
works by R. Feynman in the mid-Fifties\cite{FEY}. The American
physicist, judging virtually impossible the task of calculating
exactly the wave-function of low energy states, ingeniously
elaborated an \emph{ansatz} to guess their formal structure, on
the base of the boson symmetry property. In such a way, not only
ODLRO criterion was shown to be satisfied by helium\cite{ONPE},
but, moreover, Feynman and Cohen\cite{FEYCOE} were able to
calculate a low energy spectrum very similar to
Landau's.\footnote{As a matter of fact, Onsager and Penrose found
that about only one-tenth of the helium atoms are in the
condensate at zero temperature, even though the whole liquid is
superfluid. That means that exists a subtle relation between the
superfluid density and the condensate fraction, and sounds as
justification for Landau's opinions.}

In superconductivity, the march to the microscopic theory was
successful in 1957, with celebrated BCS theory\cite{BCS}. That is
a very complex theory, but the main ideas involved, such as the
role of the energy-gap due to some coupling between the electrons
and the relevance of low energy spectrum of excitations, were
already in London's mind since the end of the Thirties, as well as
in the Landau's description of a quantum fluid in 1941.

Indeed, the microscopic explanation of superconductive phase was
made showing that the so called Cooper pairs (electrons
interacting through the mediation of phonons in the lattice) can
form a stable state of lower energy than electrons in normal
phase\footnote{The ground state of BCS theory can be associated
with the boson Cooper pairs condensation. However, the phenomenon
of BEC in BCS was not much emphasized in the early years and
became totally accepted only in the Eighties\cite{GRI}.}. At low
temperatures, no excitations were found in the energy spectrum
from the ground state with vanishing energy (energy gap). That
explained the superconductive behavior, through the Meissner
effect. Because of the gap vanishing at a critical point, the
continuous transition resulted\footnote{In Gor'kov derivation of
Ginzburg-Landau theory from BCS, the energy gap was found to be
proportional to the order parameter $\Psi$ \cite{GOR}.}.

At the end of Fifties, the time was right for the SSB passage to
the physics of elementary interactions. In 1958, N. Bogolubov
reformulated the BCS theory building the ground state and the low
energy excitation states through the coherent superposition of
electron and hole wave-functions (Bogolubov's
\emph{quasiparticles}\cite{BOGO}). In 1961, after Heisenberg's
works about the spontaneous breaking of isotopic spin symmetry in
a nonlinear theory of elementary particles \cite{HEI}, Y. Nambu
proposed an SSB mechanism to generate nucleons masses from a
massless bare fermion theory\cite{NAM}, at the same way the energy
gap arises in BCS theory. The crucial point was the strict analogy
between the Bogolubov equations of excited states and the Dirac's
equation of quantum electrodynamics. In this analogy, a
correspondence was found between the spontaneous breaking of
charge symmetry in Bogolubov's ground state and the violation of
chirality symmetry in particle theory; the energy-gap and the
observed nucleon mass; the collective excitations of
quasiparticles and the bound nucleon pairs, or mesons.

A new way of thinking was beginning, in theoretical physics.

\section{Long live the phenomenology!}

As a matter of fact, SSB concept was elaborated through a combined
action of first principle and phenomenological physics, in a
dialectical relationship that can be traced back to that of the
late nineteenth-century between thermodynamics and kinetic
theory\cite{COHO}. In the present reconstruction, it has been
showed the role of the phenomenological approach: it was decisive
to \emph{introduce} the main ideas involved in SSB. Let us now
briefly summarize in some points the chief features which
characterized that way of making physics.

1) Accepting the Kuhnian repartition of science historical
periods, it can be said that the phenomenological approach
developed in a \emph{normal} phase of physics history. Its natural
background was the physics of problems, rather then the physics of
principles; 2) A typical frame of mind was the neglectfulness of
philosophical questions, following the Landau's remark: ''Think
less about foundations!''\cite{KAHA}; 3) The phenomenological
approach was mainly based on a \emph{bottom-up} method, that is to
say, the elaboration of a theoretical model starting from the
empirical manifestations of phenomena, in order to provide a
structural synthesis of its fundamental relations \cite{SHO}; 4)
The leading way to reach this result was to create flexible
mathematical structures, in which parameters could be fixed
according to experimental results; 5) The aim was that of looking
at the state of things, as far as possible, neglecting the
microscopic details. In this way, mathematical difficulties in
treating elementary interactions were bypassed in favor of a
\emph{universal} description; 6) In this framework, a capital
relevance was acquired by Statistical Mechanics, inasmuch it
allows to approach complex problems with the maximum of
generality; 6) From the point of view of the research
organization, the \emph{physics of problems} was particularly
cultivated in the context of schools (Sommerfeld in M\"{u}nich,
Heisenberg in Leipzig, Landau in Kharkov, K. Onnes in Leiden),
marked by common features of informality, broad interests,
attractive to young people, links with experimentalists. In these
communities, a major emphasis was given to didactics, as a
decisive instrument to train a generation of physicists in the new
techniques (it is hardly necessary to remind the monumental
Landau-Lifshitz course\cite{KHL}); 7) Ultimately, it can be said
that phenomenology not only oriented the first principle physics
towards well definite theoretical targets, but also provided
general ideas to enlighten at the same time different fields of
research.

Anyway, SSB was the yield of a physics started to become very
different from what was before.

\section*{Acknowledgements}
It is a pleasure to thank Augusto Garuccio and Sebastiano
Stramaglia for fruitful discussions and comments on the
manuscript. Thanks to Rosa De Francesco for kindness and
assistance in finding material.


\begin{thebibliography}{99}

\bibitem{BBCHN}
BROWN L., BROUT R., CAO T. Y., HIGGS P. and NAMBU Y., {\it Panel
Session: Spontaneous Breaking of Symmetry}, given at 3rd
International Symposium on the History of Particle Physics,
Stanford, CA, 24-27 Jun 1992, in {\it The Rise of the Standard
Model. Particle Physics in the 1960s and 1970s}, edited by
HODDESON L., BROWN L., RIORDAN M. and DRESDEN M. (Cambridge
University Press) 1997, pp. 478-522.

\bibitem{HU}
HUANG K., {\it Statistical Mechanics} (John Wiley \& Sons, New
York) 1987.

\bibitem{A}
ANDERSON P. W., {\it Science}, \textbf{177} (1972) p. 393-396.

\bibitem{Cast}
CASTELLANI E., {\it Studies in History and Philosophy of Modern
Physics}, \textbf{33/2} (2002) pp. 251-267 [arXiv:
physics/0101039v1 [physics.hist-ph]].

\bibitem{N} NAMBU Y., {\it J. Phys. Soc.
Jap.}, \textbf{76}, 11 (2007) p. 111002, available on line
http://jpsj.ipap.jp/link?JPSJ/76/111002/pdf

\bibitem{BRE} BROUT R. and ENGLERT F., talk given at International Europhysics Conference on
High-Energy Physics (HEP 97), Jerusalem, Israel, 19-26 Aug 1997
[arXiv:hep-th/9802142v2].

\bibitem{HBTW} HODDESON L., BRAUN E., TEICHMANN J. and
WEART S., {\it Out of the Crystal Maze: Chapters from the History
of Solid State Physics} (Oxford University Press, New York) 1992.

\bibitem{HBTW1} HODDESON L., BRAUN E., TEICHMANN J. and
WEART S.\cite{HBTW}, p. 373.

\bibitem{BC} BROWN L. and CAO T. Y., {\it Historical Studies in the Physical and
Biological  Sciences}, \textbf{21} (1991) pp. 211-235.

\bibitem{R} RADICATI L., in {\it Symmetries in Physics (1600-1980): Proceedings of the 1st
International Meeting on the History of Scientific Ideas}, Sant
Feliu de Guixols, Catalogna, Spain, September 20-26, 1983, ed. by
DONCEL M., HERMANN A., MICHEL L and PAIS A. (Universitat
Aut\`{o}noma de Barcelona, Barcelona) 1987, pp. 197-206.

\bibitem{S} SHIRKOV D. V., {\it Physics Uspekhi}, \textbf{52}, 6 (2009) p. 549 [arXiv:0903.3194v2
[physics.hist-ph]].

\bibitem{JL} JONA-LASINIO, G., {\it Cross fertilization in
theoretical physics}, in {\it Symmetries in Physics: Philosophical
Reflections}, edited by BRADING K. and CASTELLANI E. (Cambridge
University Press), 2003, pp. 315-320.

\bibitem{E} EULER L., originally published in {\it Memoires de l'academie des sciences de
Berlin}, \textbf{13} (1759) pp. 252-282, reprinted in {\it
Leonhardi Euleri Opera Omnia}, ser. II, \textbf{XI}, sec. II,
(Orell F\"{u}ssli Turici, Berna) 1960, p. 345, available on line
http://www.math.dartmouth.edu/~euler/docs/originals/E238.pdf

\bibitem{J} JACOBI C. G. J., originally published in \emph{Poggendorff Annalen der
Physik und Chemie}, \textbf{33} (1834) pp. 229-233, reprinted in
\emph{Gesammelte Werke}, \textbf{2} (Reiner, Berlin) 1882-1891,
pp. 17-72, available on line
http://gallica.bnf.fr/ark:/12148/bpt6k90215d.image.f1

\bibitem{B} BACHELARD G., \emph{La formation de l'esprit scientifique}
(Librairie philosophique Vrin, Paris) 1999 (1\`{e}re \'{e}dition:
1938).

\bibitem{BG} BAKER M. and GLASHOW S. L., \emph{Phys. Rev.}, \textbf{128}, 5 (1962)
pp. 2462-2471.

\bibitem{H} HADOT P., \emph{The Veil of Isis: An Essay on the History of the Idea of Nature}
En. tr. CHASE M. (Harvard University Press) 2006.

\bibitem{C} CURIE P., \emph{J. Phys. (Paris)}, 3e s\'{e}rie (1894)
pp. 393-415.

\bibitem{BEN} B\'{E}NARD H., in \emph{Revue G\'{e}n\'{e}rale des Sciences} \textbf{11} (1900), pp. 1261-1271,
1309-1328.

\bibitem{BI} BIRKHOFF G, \emph{Hydrodynamics} (Princeton University Press) 1950.

\bibitem{SL} LIE S. and ENGEL F., \emph{Theorie der
Transformationsgruppen} (B.G. Teubner, Leipzig) 1988-1893.

\bibitem{PW} WEISS P., \emph{Journal de physique theorique et appliqu\'{e}}, \textbf{61} (1907)
pp. 661-690.

\bibitem{OZ} ORNSTEIN L. S. and ZERNIKE F., \emph{Accidental Deviations of Density and Opalescence
at the Critical Point of a Single Substance}, in
\emph{Verhandelingen der Konenlijke Akademie van Wetenschappen te
Amsterdam} \textbf{17} (1914) pp. 793-806.

\bibitem{GBJLD} GORSKY W., \emph{Z. Phys.}, \textbf{50} (1928) p.
64; BORELIUS G. JOHANSSON C. H. and LINDE J. O., \emph{Ann. Phys.
(Leipzig)}, \textbf{86} (1928) p. 291; DEHLINGER U., \emph{Z.
Phys. Chem.} \textbf{26} (1934) p. 343.

\bibitem{BW} BRAGG W. L. and Williams E. J., \emph{Proc. R. Soc. London}
\textbf{A145} (1934) p. 699; \textbf{A151} (1935) p. 540;
\textbf{A152} (1935) p. 231.

\bibitem{BE} BETHE H. A. Bethe, \emph{Proc. R. Soc. London}, \textbf{A150} (1935) p.
552-575.

\bibitem{PA} PEIERLS R., \emph{Proc. R. Soc. London}, \textbf{A 154} (1936) p.
207-222.

\bibitem{KI} KIRKWOOD J. G., \emph{J. Chem. Phys.}, \textbf{6} (1938) p. 70.

\bibitem{GC} GORTER C. J. and CASIMIR H. B. G., \emph{Physica}, \textbf{1} (1934) p. 306-320.

\bibitem{GOR} GORTER C.J.G. Gorter, \emph{Reviews of Mod. Phys.} \textbf{36} (1964) p. 3-7 .

\bibitem{JA} JAEGER G., \emph{Archive for History of Exact Sciences}, \textbf{53}, 1
(1998) pp. 51-81.

\bibitem{LA1} LANDAU L., \emph{Phys. Z. Sowjetunion}, \textbf{11},
(1937) I part p. 26 , II part p. 545; Engl. tr. in \emph{Collected
Papers of L. D. Landau}, (Pergamon, Oxford) 1965, p. 193.

\bibitem{PO} POKROVSKY V. L., \emph{History of Physics Newsletter} (APS) \textbf{7}, 3
(August 1998) available online at
http://www.aps.org/FHP/pokrovsky.html.

\bibitem{ST} STANLEY H. E., \emph{Phase transitions and critical
phenomena}(Clarendon Press, Oxford) 1972, p. 168.

\bibitem{JDB} de BOER J. \emph{Physica}, \textbf{73} (1974) p.
1-27.

\bibitem{AN} ANDERSON P. W., \emph{Basic notions of Condensed Matter} (The Benjamin/Cummings Publishing
Company, Menlo Park, California) 1984, p. 19.

\bibitem{GO} GOLDSTONE J., \emph{Il Nuovo Cimento}, \textbf{XIX}, 1
(1961) p. 154.

\bibitem{BL} BLOCH F., \emph{Z. Phys.}, \textbf{61} (1930) p.
206.

\bibitem{KO1} KOJEVNIKOV A. B., \emph{Stalin's Great Science.
The Times and Adventures of Soviet Physicist.} (Imperial College
Press, London) 2004.

\bibitem{KO2} KOJEVNIKOV A. B., \emph{Historical Studies in the Physical and Biological
Sciences} \textbf{33}, 1 (2002) pp. 161–192 available on line:
http://www.history.ubc.ca/documents/Collectivism.pdf

\bibitem{LO} LONDON F., \emph{Phys. Rev.} \textbf{54 }(1938) pp.
947-954.

\bibitem{LA2} LANDAU L., \emph{Zh. Eksp. Teor. Fiz.}, \textbf{11} (1941) p. 592;
\emph{J. Phys. USSR}, \textbf{5}, (1941) p. 71 ; Engl. tr. in
\emph{Collected Papers of L. D. Landau}, (Pergamon, Oxford) 1965,
p. 301.

\bibitem{BOG} BOGOLUBOV N.N., \emph{J. Phys. USSR}, \textbf{11} (1947) pp. 23-32.

\bibitem{LOLO} LONDON F. and LONDON H., \emph{Proc. R. Soc. London}, \textbf{A149} (1935) p.
71.

\bibitem{GILA} LANDAU L. and GINZBURG V.,\emph{Zh. Eksp. Teor. Fiz.}, \textbf{20} (1950) p.
1064, Engl. tr. in \emph{Collected Papers of L. D. Landau},
(Pergamon, Oxford) 1965, p. 546.

\bibitem{GRI} GRIFFIN A., \emph{A brief history of our understanding of BEC: from Bose to Beliaev},
Lecture given at the BEC Varenna Summer School, July 7-17, 1998,
in \emph{Bose-Einstein Condensation in Atomic Gases}, edited by
INGUSCIO M., STRINGARI S. and WIEMAN C. (Italian Physical Society)
1999 [arXiv:cond-mat/9901123v1].

\bibitem{GORK} GOR'KOV L. P. \emph{Sov. Phys. JETP}, \textbf{36}, 6 (1959), p. 1364.

\bibitem{ARKR} ARANSON I. S. and KRAMER L., \emph{Rev. Mod. Phys.}, \textbf{74} (2002) p.
99-143.

\bibitem{PEN} PENROSE O., \emph{Phil. Mag.}, \textbf{42} (1951) p. 1373.

\bibitem{ONPE} PENROSE O. and ONSAGER L., \emph{Phys. Rev.}, \textbf{104} (1956) p.
576.

\bibitem{YA} YANG C. N., \emph{Rev. Mod. Phys.}, \textbf{34} (1962) p.
694-704.

\bibitem{FEY} FEYNMAN R. P., \emph{Phys. Rev.}, \textbf{91}
(1953) p. 1291; \textbf{91} (1953) p. 1301;  \textbf{94} (1954) p.
262.

\bibitem{FEYCOE} FEYNMAN R. P. and COHEN M., \emph{Phys. Rev}, \textbf{102}, 5 (1956) p.
1189.

\bibitem{BCS} BARDEEN J., COOPER L. N. and SCHRIEFFER J. R., \emph{Phys. Rev.}, \textbf{108} (1957)
p. 1175.

\bibitem{BOGO} BOGOLUBOV N. N., \emph{Nuovo Cimento},
\textbf{7} (1958) p. 794-805.

\bibitem{HEI} von D\"{U}RR H.P., HEISENBERG W., MITTER H., SCHLIEDER S. and YAMAZAKI K.,
\emph{Z. Naturforsch.}, \textbf{14 a} (1959) p. 441-485.

\bibitem{NAM} NAMBU Y. and JONA-LASINIO G., \emph{Phys. Rev.}, I part \textbf{122}, 1 (1961) p.
345; II part \textbf{124}, 1 (1961) p. 246.

\bibitem{COHO} CLARK P., \emph{Atomism versus Thermodynamics}, in \emph{Method and
Appraisal in the Physical Science}, edited by HOWSON C. (Cambridge
University Press) 1976, p. 41.

\bibitem{KAHA} HALL K. \emph{Think less about foundations: a short course on the
course of theoretical physics of Landau and Lifshitz}, in
\emph{Pedagogy and the practice of science : historical and
contemporary perspectives}, ed. by KAISER D. (MIT Press,
Cambridge, Mass) 2005, p. 253-286.

\bibitem{SHO} SHOMAR T. L. \emph{Current science}, \textbf{94}, 10 (25 May
2008) p. 1256, available on line:
http://www.ias.ac.in/currsci/may252008/1256.pdf

\bibitem{KHL} HALL K., \emph{The schooling of Lev Landau}, in \emph{Osiris}, \textbf{23}
(The University of Chicago Press) edited by GORDIN D., HALL K. and
KOJEVNIKOV A., 2008, p. 230.

\bibitem{GOE} GOETHE J. W., \emph{Epirrhema}, in \emph{Goethes
Werke}, ed. by COTTA J. G. (Stuttgart und T\"{u}bingen) 1827-1842
p. 86. Cited and translated in\cite{H}, p. 256.

\bibitem{ONS} ONSAGER L., \emph{Phys. Rev.}, \textbf{65} (1944) pp.
117-149.


\end{thebibliography}
\end{document}